\pdfoutput=1
\documentclass[10pt,journal,a4paper,twoside]{IEEEtran}

\usepackage[latin2]{inputenc}
\usepackage[T1]{fontenc}
\usepackage{graphicx}
\usepackage{amsfonts}
\usepackage{amsmath}
\usepackage{amssymb}
\usepackage[hang]{subfigure}
\usepackage{listings}
\usepackage{array}
\usepackage{fullpage}

\lstdefinestyle{nonumbers}
{numbers=none}

\lstdefinelanguage{qpc}
  {keywords={for, to, if, then, while, do,else,and,or,qbit, qnibble,elsif,qreg},
    sensitive=true,
    comment=[l]{C:},
    morestring=[b]"
  }

\newcommand{\bra}[1]{\langle{#1}|}
\newcommand{\ket}[1]{|{#1}\rangle} 
\newcommand{\href}[1]{}
\newcommand{\HH}{\mathcal{H}}
\newcommand{\bt}{\ttfamily}

\newcommand{\1}{\mbox{$\mathbb{I}$}}

\newcommand{\ketbra}[2]{\ket{#1}\bra{#2}}

\newcommand{\Tr}[1]{\mathrm{Tr}\left(#1\right)}
\newcommand{\PTr}[2]{\mathrm{Tr}_{#1}\left(#2\right)}
\newcommand{\qo}{\texttt{quan\-tum-octa\-ve}}

\newcommand{\qvalueof}[2]{[[#1]]_{#2}}

\newcommand{\eg}{\emph{eg.}}
\newcommand{\ie}{\emph{ie.}}

\title{Extending scientific computing system with structural quantum programming
capabilities}

\author{Piotr Gawron\thanks{E-mail address: gawron@iitis.pl (Corresponding
author)} \qquad Jerzy Klamka \qquad Jaros\l{}aw Adam Miszczak \qquad Ryszard
Winiarczyk\\
\mbox{} \\ %
The Institute of Theoretical and Applied Informatics\\
of the Polish Academy of Sciences\\ 
Ba\l{}tycka 5, 44-100 Gliwice, Poland
}

\IEEEspecialpapernotice{November 12, 2009}

\begin{document}
\maketitle

\begin{abstract}
We present a basic high-level structures used for developing quantum programming
languages. The presented structures are commonly used in many existing quantum
programming languages and we use quantum pseudo-code based on QCL quantum
programming language to describe them.

We also present the implementation of introduced structures in GNU Octave
language for scientific computing. Procedures used in the implementation are
available as a package \qo, providing a library of functions, which facilitates
the simulation of quantum computing. This package allows also to incorporate
high-level programming concepts into the simulation in GNU Octave and Matlab. As
such it connects features unique for high-level quantum programming languages,
with the full palette of efficient computational routines commonly available in
modern scientific computing systems.

To present the major features of the described package we provide the
implementation of selected quantum algorithms. We also show how quantum errors
can be taken into account during the simulation of quantum algorithms using \qo\
package. This is possible thanks to the ability to operate on density matrices.
\end{abstract}


\section{Introduction}
Quantum information theory aims to harness the quantum nature of information
carriers in order to develop more efficient algorithms and more secure
communication protocols~\cite{NC2000, hirvensalo, BKW2001a, BKW2001b}.
Unfortunately, counterintuitive laws of quantum mechanics make the development
of new quantum information processing procedures highly non-trivial task. This
can be seen as one of the reasons why only few truly quantum algorithms were
proposed~\cite{shor03why, shor04progress}.

The laws of quantum mechanics are in many cases very different from those we
know from the classical world. That is why one needs to seek for the novel
methods for describing information processing which involves quantum elements.
To this day several formal models were proposed for the description of quantum
computation process~\cite{Deutsch1985,Deutsch1989, bettelli03towards,
Gudder2003, Vantonder2004, Moore2000275}.

The most popular of them is the quantum circuit model \cite{Deutsch1985}, which
is tightly connected to the physical operations implemented in a laboratory. It
allows to operate with the basic ingredients of quantum information processing
-- namely qubits, unitary evolution operators and measurements. However, it does
not provide too much flexibility concerning operations on more sophisticated
elements required to develop scalable algorithms and protocols \eg\ quantum
registers or classical controlling operations.

Another model widely used in the study of theoretical aspects of quantum
information processing is the quantum Turing machine~\cite{Deutsch1985}. This
model is mainly used in the analysis of quantum complexity
problems~\cite{bernstein97complexity}. Its main advantage it that it provides
method of comparing efficiency of classical and quantum algorithms.
Unfortunately quantum Turing machine, in analogy to its classical counterpart,
operates on very basic data and thus it cannot be easily used to construct
quantum algorithms.

Both quantum circuit model and quantum Turing machine share some serious
drawbacks concerning lack of support for high-level programming and very limited
flexibility. These problems were addressed in the recent research in the area of
quantum programming languages~\cite{gay05quantum, MiszczakPHD, Gay2008}.

Quantum programming languages are based on the Quantum Random Access Machine
(QRAM) model. QRAM is equivalent, with respect to its computational power, to
the quantum circuit model or quantum Turing machine. However, it has strictly
distinguished two parts: quantum and classical. The quantum part is responsible
for performing parts of a algorithm which cannot be computed efficient by a
classical machine. The classical part is used to control quantum computation.
This model is used as a basis for most quantum programming
languages~\cite{gay05quantum}. 

Among high-level programming languages designed for quantum computers we can
distinguish imperative and functional languages. The later are seen by many
researchers as a means of providing robust and scalable methods for developing
quantum algorithms~\cite{altenkirch05functional}. We, however, focus on the
imperative paradigm as it provides more straightforward way of implementing
high-level quantum programming concepts.

This paper is organized as follows. In Section \ref{sec:qram} we briefly
describe the QRAM model of quantum computer and introduce the quantum
pseudocode, which was designed to describe this model. In
Section~\ref{sec:basic} we introduce high-level programming structures used in
quantum programming languages. These structure are based on the QRAM model of
quantum computer. In Section~\ref{sec:qo} the implementation of presented
concepts is described and \qo\ package is presented. In Section~\ref{sec:apps}
implementation of quantum algorithms using \qo\ package is presented. Also the
analysis of quantum errors is provided in the case of quantum search algorithm.
Finally Section~\ref{sec:summary} summarize the presented work and provides
reader with some concluding remarks.

\section{QRAM model of quantum computation}\label{sec:qram}
Quantum random access machine is interesting for us since it provides convenient
model for developing quantum programming languages. However, these languages and
basic concepts used to develop them are our main area of interest. For this
reason here we provide only the very basic introduction to the QRAM model.
Detailed description of this model is given in \cite{knill96conventions} and
\cite{Oemer2003} together with the description of hybrid architecture used in
quantum programming.

\subsection{Classical RAM model}
As in many situations in quantum information science, the QRAM models is based
on the concepts developed to describe classical computational process -- in this
case on the Random Access Machine (RAM) model. The classical model of random
access machine (RAM) is the example of more general register machines
\cite{cook73time, papadimitriou, shepherdson63computability}. 

The Random Access Machine consists of an unbounded sequence of memory registers
and finite number of arithmetic registers. Each register may hold an arbitrary
integer number. The programme for the RAM is a finite sequence of instructions
$\Pi=(\pi_1,\ldots,\pi_n)$. At each step of execution register $i$ holds an
integer $r_i$ and the machine executes instruction $\pi_\kappa$, where $\kappa$
is the value of the programme counter. Arithmetic operations are allowed to
compute the address of a memory register.

Despite the difference in the construction between Turing machine and RAM, it
can be easily shown that Turing machine can simulate any RAM machine with
polynomial slow-down only~\cite{papadimitriou}. The main advantage of the RAM
models is its resemblance with the real-world computers.

\subsection{Quantum RAM model and quantum pseudocode}
\begin{figure}[ht]
 \centering
 \includegraphics[width=\columnwidth]{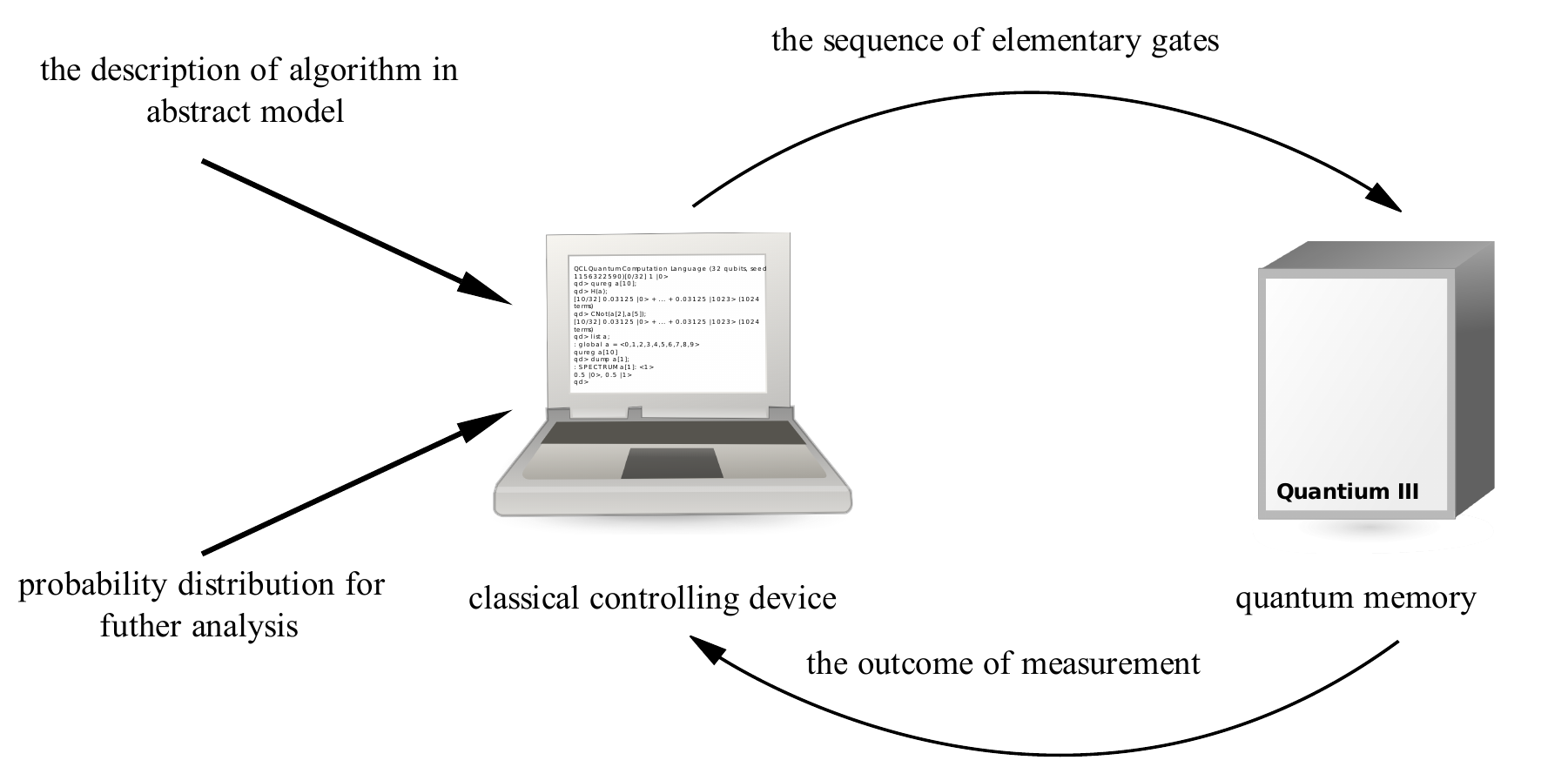}
 \caption[Quantum random machine model]{The model of classically controlled
quantum machine \cite{Oemer2003}. Classical computer is responsible for
performing unitary operations on quantum memory. The results of quantum computation
are received in the form of measurement results.}
 \label{fig:qram}
\end{figure}

Quantum Random Access Machine (QRAM) model is build as an extension of the
classical RAM model. Its main goal is to provide the ability to exploit quantum
resources. Moreover, it can be used to perform any kind of classical computation.
The QRAM allows us to control operations performed on quantum registers and
provides the set of instructions for defining them. Schematic presentation of
such architecture is provided in Figure~\ref{fig:qram}.

The quantum part of QRAM model is used to generate probability distribution.
This is achieved by performing measurement on quantum registers. The obtained
probability distribution has to be analysed using a classical computer.

Quantum algorithms are, in most of the cases, described using the mixture of
quantum gates, mathematical formulas and classical algorithms. The first attempt
to provide a uniform method of describing quantum algorithms was made in
\cite{cleve96schumacher}, where the author introduces a high-level notation
based on the notation known from computer science textbooks \cite{cormen,
hopcrof03wprowadzenie}.

In \cite{knill96conventions} Knill introduced the first formalized language for
description of quantum algorithms. Moreover, it was tightly connected with the
model of Quantum Random Access Machine.

Quantum pseudocode proposed by Knill \cite{knill96conventions} is based on
conventions for classical pseudocode proposed in \cite[Chapter 1]{cormen}.
Classical pseudocode was designed to be readable by professional programmers, as
well as people who had done a little programming. Quantum pseudocode introduces
operations on quantum registers. It also allows to distinguish between classical
and quantum registers. In quantum pseudocode quantum registers are distinguished
with an underline. They can be introduced by applying quantum operations to
classical registers or by calling a subroutine which returns a quantum state. In
order to convert a quantum register into a classical register measurement
operation has to be performed.

The example of quantum pseudocode is presented in Listing \ref{lst:qpc-example}.
It shows the main advantage of QRAM model over quantum circuits model -- the
ability to incorporate classical control into the description of quantum
algorithm.

\begin{figure}[ht]
  \begin{center}
	\includegraphics[scale=1.2]{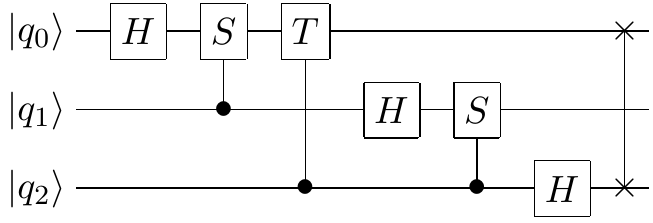}
  \end{center}
  \caption[Quantum Fourier transform for three qubits]{Quantum circuit
representing quantum Fourier transform for three qubits. Elementary gates used
in this circuit are described in \cite{NC2000}.}
  \label{fig:qft3}
\end{figure}

\begin{lstlisting}[language=qpc, float=ht, mathescape=true, escapechar=\%,
label=lst:qpc-example,caption={[Quantum pseudocode for quantum Fourier
transform]Quantum pseudoceode for quantum Fourier transform on $d$ qubits.
Quantum circuit for this operation with $d=3$ is presented in
Figure~\ref{fig:qft3}.}]
%{\bf Procedure:} {\sc Fourier}($\underline{a},d$)\\
{\bf Input:} A quantum register $\underline{a}$ with $d$ qubits. Qubits are
numbered form 0 to $d-1$.\\
{\bf Output:} The amplitudes of $\underline{a}$ are Fourier transformed over
$\ \Z_{2^d}$.\\
%
C: assign value to classical variable
$\omega \leftarrow e^{i2\pi/2^d}$
C: perform sequence of gates
for  $i = d-1$ to $i=0$
  for $j = d-1$ to $j=i+1$
  %\underline{\bf if}% $\underline{a_j}$ %\underline{\bf
then}% $\mathcal{R}_{\omega^{2^{d-i-1+j}}}(\underline{a_i})$
   C: number of loops executing phase depends on
   C: the required accuracy of the procedure
  $\mathcal{H}(\underline{a_i})$

C: change the order of qubits
for $j = 0$ to $j=\frac{d}{2}-1$
  $\mathcal{SWAP}(\underline{a_j},\underline{a_{d-a-j}})$
\end{lstlisting}

Operation $\mathcal{H}(\underline{a_i})$ executes a quantum Hadamard gate on a
quantum register $\underline{a_i}$ and $\mathcal{SWAP}(\underline{a_i},
\underline{a_j})$ performs SWAP gate between $\underline{a_i}$ and
$\underline{a_j}$. Operation $\mathcal{R}_{\phi}(\underline{a_i})$ executes a
quantum gate $R(\phi)$ is defined as
\begin{equation}
 R(\phi) =     \left(
    \begin{array}{cc}
      1 & 0 \\
      0 & e^{i\phi} \\
    \end{array} 
    \right),
\end{equation}
on the quantum register $\underline{a_i}$. Using conditional construction
\begin{lstlisting}[language=qpc,mathescape=true,escapechar=\%]
 %\underline{\bf if}% $\underline{a_j}$ %\underline{\bf
then}% $\mathcal{R}_{\phi}(\underline{a_i})$
\end{lstlisting}
it is easy to define controlled phase shift gate. Similar construction exists in
QCL quantum programming language~\cite{Oemer2003}. In Section~\ref{sec:qo} we 
describe implementation of this construction in \qo.

The measurement of a quantum register can be indicated using an assignment
\begin{lstlisting}[language=qpc,mathescape=true,escapechar=\%]
 $a_j \leftarrow \underline{a_j}$.
\end{lstlisting}

\subsection{Requirements for quantum programming language}
Taking into account QRAM model we can formulate basic requirements which have to
be fulfilled by any quantum programming language \cite{bettelli03towards,
bettelliPHD, MiszczakPHD}.

\begin{itemize}
  \item{{\bf Completeness:}} Language must allow to express any quantum circuit 
and thus enable the programmer to code every valid quantum programme written as
a quantum circuit. \index{quantum!circuit}
  \item{{\bf Extensibility:}} Language must include, as its subset, the language
implementing some high level classical computing paradigm. This is important
since some parts of quantum algorithms (for example Shor's algorithm) require
non-trivial classical computation.
  \item{{\bf Separability:}} Quantum and classical parts of the language should
be separated. This allows to execute any classical computation on purely
classical machine without using any quantum resources.
  \item{{\bf Expressivity:}} Language has to provide high level elements for
facilitating the quantum algorithms coding.
  \item{{\bf Independence:}} The language must be independent from any
particular physical implementation of a quantum machine. It should be possible
to compile a given programme for different architectures without introducing any
changes in its source code.
\end{itemize}

\section{High-level programming structures}\label{sec:basic}

\subsection{Quantum memory}
Quantum memory is a set of qubits indexed by integer numbers. Quantum register
is a set of indices pointing to distinct qubits. We will denote those registers
as $r_1, r_2, \ldots$ or in case of single qubits as $q_1, q_2, \ldots$. The
state of a quantum memory is a quantum state of size equal to the number of
qubits. In the case of \qo\ we operate on density matrices (although some
operations on state vectors are allowed). We will denote the state of the
quantum memory by $\rho$.

Following operations on quantum memory are allowed:
\begin{itemize}
	\item Allocation of new register of size $n$: 
	\begin{equation}\label{equ:extentionevolution}
		\rho_{t+1}=\rho_{t}\otimes\ketbra{\underbrace{0\ldots 0}_n}{\underbrace{0\ldots 0}_n},
	\end{equation}
	where $\otimes$ denotes tensor product, $\ket{\cdot}$ the~column vector and $\bra{\cdot}$ the~dual vector.
	\item Deallocation of a register indexed by register $r$:
	\begin{equation}\label{equ:traceevolution}
		\rho_{t+1}=\PTr{r}{\rho_{t}},
	\end{equation}
	where $\PTr{r}{\rho}$ denotes partial trace of $\rho$ with regard to the subsystem indexed by~$r$.
	\item Unitary evolution $U$ of the quantum memory: 
	\begin{equation}\label{equ:unitaryevolution}
		\rho_{t+1}=U\rho_{t} U^\dagger.
	\end{equation}
	\item Application of quantum channel $K_i$ on the quantum memory: 
	\begin{equation}\label{equ:krausevolution}
		\rho_{t+1}=\sum_{i} K_i\rho_{t} {K_i}^\dagger.
	\end{equation}
	\item Measurement in the computational basis:
	\begin{align}\label{equ:measurment}
		\rho_{t+1}=\sum_{i} \ketbra{i}{i}\rho_{t} \ketbra{i}{i}, \\
		P(i)=\Tr{\ketbra{i}{i}\rho_t},
	\end{align}
	where $i$ enumerates the states of computational basis.
\end{itemize}

For a solid introduction to quantum computation the reader may refer to book by
Nielsen and Chuang \cite{NC2000}, where all the needed notions are explained in
detail.

In quantum computation, construction of the unitary gate is the essential part
of quantum algorithm (program) design process. It is a~difficult task to write
a~quantum program using only elementary set of gates \ie\ CNot and one qubit
rotations. Therefore it is desirable to introduce some techniques that
facilitate the process of composition of complex quantum gates. Some of those
techniques are presented below. We will refer to implementation of those
techniques in \qo\ which is described in details in section~\ref{sec:qo}.

\subsection{Composed and controlled gates}

\subsubsection{Composed gate}
Given one-qubit unitary gate $G$, quantum register $r$, and size of the gate $s$
we can construct composed gate $U_r^s$ according to the formula:
\begin{equation}\label{equ:productgage}
	U_r^s=\bigotimes_{i=1}^s X_i, \text{where }
	X_i=
	\left\lbrace
	\begin{array}{ccc}
	 G&\text{if}& i\in r, \\
	 \1&\text{if}& i\notin r
	\end{array}\right..
\end{equation}

\subsubsection{Controlled gate with multiple controls}

Given one-qubit unitary gate $G$, quantum register $r_c$ we call control, and
quantum register $r_t$ we call target, and size of the gate $s$ we can construct
controlled gate $U_{r_t|r_c}^s$ according to the formula:
\begin{equation}\label{equ:controlledgate}
\begin{array}{rcl}
U_{r_t|r_c}^s&=&\bigotimes_{i=1}^s X_i+\bigotimes_{i=1}^s Y_i, \quad\text{where }
\\
X_i&=&
\left\lbrace
\begin{array}{ccl}
\ketbra{0}{0}&\text{if}& i\in r_c, \\
\1&\text{if}& i\notin r_c,
\end{array}\right.
\\
Y_i&=&
\left\lbrace
\begin{array}{ccl}
G&\text{if}& i\in r_t, \\
\ketbra{1}{1}&\text{if}& i\in r_c, \\
\1&\text{if}& i\notin r_c\cup r_t
\end{array}\right.
\end{array}.
\end{equation}
We assume that $r_t\cap r_c =\varnothing$. Sometimes we will omit the size
parameter~$s$.

\subsection{Conditionals}
One of high-level technique used in quantum programming are quantum
conditions~\cite{GawronPHD}. The main idea behind quantum conditions is
construction of quantum gates controlled by predicates on control registers.

\subsubsection{Condition on quantum variable}

The if-then-else structure controlled by a quantum variable and acting on a
quantum variable was introduced in QCL \cite{Oemer2000, Oemer2003}.

\begin{figure}[htp]
\centering
\begin{tabular}{c|m{.3\linewidth}}
\begin{lstlisting}[language=qpc, mathescape=true]{}
qbit $q_1, q_2$

if ($q_1$) then
  $\mathcal{G}_1(q_2$)
else
  $\mathcal{G}_2(q_2$)
\end{lstlisting}
&
\includegraphics[width=0.4\columnwidth]{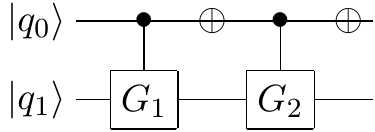}\\
\end{tabular}
\caption{Example of a simple quantum if-then-else structure.}
\label{fig:simple-conditional}
\end{figure}

In Figure~\ref{fig:simple-conditional} the realisation of this concept is
presented. If qubit $q_0$ is in the state $\ket{1}$ the $G_1$ gate is applied to
qubit $q_1$, otherwise the gate $G_2$ is applied.

We may write this circuit in the following way: 
\begin{equation}
	IF_{q_0}(G_1{q_1})ELSE(G_2{q_1})=Not_{q_0}G_{2{q_1|q_0}}Not_{q_0}G_{1{q_1|q_0}}.
\end{equation}

For a~given control register $r_c$, target register $r_t$ and two quantum gates
$G_1$ and $G_2$, we may define quantum condition in the more general way, namely
\begin{eqnarray}\label{equ:quantumif}
	IF_{r_c}(G_1{r_t})ELSE(G_2{r_t})=\\ \nonumber \prod_{i\in\mathcal{P}(r_t)\setminus \{\varnothing\}} \left(Not_{i}G_{2{r_t|r_c}}Not_{i}\right) G_{1{r_t|r_c}},
\end{eqnarray}
where $\mathcal{P}(\cdot)$ denotes the power set.

\subsubsection{Condition on mixture of classical and quantum variables}

One may consider the relation between the state of quantum register and the
value of the classical variable. By $\qvalueof{x}{r}$ we will denote a numerical
value of ordered (in ascending order) elements of the register $x$ in regard to
register $r$, for example the value of $\qvalueof{\{4,9\}}{\{2,4,7,9\}}$ is 10.
By $[r]$ we will denote the value of the register in order to use it as argument
for arithmetic comparison. For example $[r]<4$ means: ``all those values of $r$
that are less than four.''

\begin{figure}[htp]
 \centering
\begin{tabular}{|l|l|}
\hline
\multicolumn{2}{|c|}{Less than} \\
\hline
pseudocode & \qo\ \\
\hline
\begin{lstlisting}[language=qpc, mathescape=true]{}
qnibble $r$
qbit $q_1, q_2$

if ($r<4$) then
  $\mathcal{G}_1(q_2)$
else
  $\mathcal{G}_2(q_1)$
\end{lstlisting}
&
\begin{lstlisting}[language=octave]{}
r=newregister(4);
q1=newregister(1);
q2=newregister(1);

qif(...
	qrlt(qureg(q1),4),...
	{G1,qureg(q2)}, ...
	{G2,qureg(q1)})
\end{lstlisting}\\
\hline
\multicolumn{2}{|c|}{
Quantum Circuit
}\\
\hline
\multicolumn{2}{|c|}{
	\includegraphics[width=0.6\columnwidth]{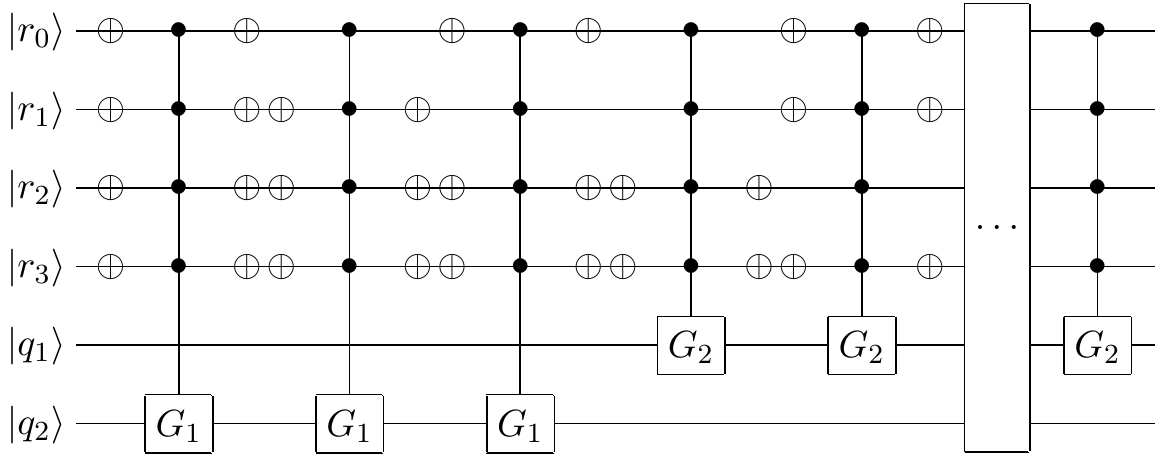}
}\\
\hline
\end{tabular}
 \caption{Example of quantum conditional operation with inequality.}
 \label{fig:condition-lt}
\end{figure}

Code and circuit in Figure~\ref{fig:condition-lt} show the idea and the
implementation of conditional operation controlled by expression `less than'
operating on classical constant and quantum register.

In the general case, gate implementing any relation (marked as $\circledast$)
can be constructed in the following~way:
\begin{eqnarray}\label{equ:quantumif-relation}
IF_{[r_c]\circledast N}(G_1{r_t})ELSE(G_2{r_t})=&&\\ \nonumber
	\prod_{i\in\mathcal{F}} \left(Not_{i}G_{2{r_t|r_c}}Not_{i}\right) 
	\prod_{i\in\mathcal{T}} \left(Not_{i}G_{1{r_t|r_c}}Not_{i}\right),
\end{eqnarray}
where sets $\mathcal{T}$ and $\mathcal{F}$ are defined as follows:
\begin{align}
\mathcal{T}=\mathcal{P}(r_c)\setminus\{x|x\in \mathcal{P}(r_{c}) \wedge \qvalueof{x}{r_{c}}\circledast N\}, \\
\mathcal{F}=\mathcal{P}(r_c)\setminus\{x|x\in \mathcal{P}(r_{c})\wedge \neg(\qvalueof{x}{r_{c}}\circledast N)\}.
\end{align}
Note that $\mathcal{T}\cup\mathcal{F}=\mathcal{P}(r_c).$

In \qo\ standard arithmetic relations $=, \neq, <, >, \leq, \geq$ are
implemented.
\subsection{Expressions}
We may consider more complicated expression on quantum registers. For example
logical operators and quantum pointers. Logical operators allow to apply an
controlled operation to the target register only if a~given logical expression
on control registers is true. A~quantum pointer allows to apply controlled gate
on the target register selected by the state of the control register.

\subsubsection{Logical expressions on quantum variables}
The gate that implements logical expression (denoted here by $\diamond$) is
constructed according to the following equation:
\begin{align}\label{equ:quantumif-expression}
IF_{[r_{c_1}]\circledast_1 N_1 \diamond [r_{c_2}]\circledast_2 N_2}(G_1{r_t})ELSE(G_2{r_t})= \nonumber \\
	= \prod_{i\in\mathcal{F}} \left(Not_{i}G_{2{r_t|r_c}}Not_{i}\right) 
	\prod_{i\in\mathcal{T}} \left(Not_{i}G_{1{r_t|r_c}}Not_{i}\right),
\end{align}
where sets $\mathcal{T}$ and $\mathcal{F}$ are defined as follows:
\begin{align}
\mathcal{T}=\mathcal{P}(r_c)\setminus\{x_1\cup x_2|x_1\subseteq r_{c_1}, x_2\subseteq r_{c_2} \\ \nonumber \wedge \left(\qvalueof{x_1}{r_{c_1}}\circledast_1 N_1 \diamond \qvalueof{x_2}{r_{c_2}}\circledast_2 N_2\right)\}, \\
\mathcal{F}=\mathcal{P}(r_c)\setminus\{x_1\cup x_2|x_1\subseteq r_{c_1}, x_2\subseteq r_{c_2} \\ \nonumber \wedge \neg \left(\qvalueof{x_1}{r_{c_1}}\circledast_1 N_1 \diamond \qvalueof{x_2}{r_{c_2}}\circledast_2 N_2\right)\}
\end{align}
and $r_c=r_{c_1}\cup r_{c_2}$.

\begin{figure}[htp]
 \centering
\begin{tabular}{|l|l|}
\hline
\multicolumn{2}{|c|}{And} \\
\hline
pseudocode & {\qo} \\
\hline
\begin{lstlisting}[language=qpc,mathescape=true]{}
qbit $q_1, q_2, q_3, q_4$

if ($q_1$ and $q_2$) then
  $\mathcal{G}_1(q_3)$
else
  $\mathcal{G}_2(q_4)$
\end{lstlisting}
&
\begin{lstlisting}[language=octave]{}
q1=newregister(1);
q2=newregister(1);
q3=newregister(1);
q4=newregister(1);

qif(...
	qrand(...
		qreq(qureg(q1),1),...
		qreq(qureg(q2),1) ),..
	{G1,qureg(q3)}, ...
	{G2,qureg(q4)})
\end{lstlisting}\\
\hline
\multicolumn{2}{|c|}{
Quantum Circuit
}\\
\hline
\multicolumn{2}{|c|}{
	\includegraphics[width=0.8\columnwidth]{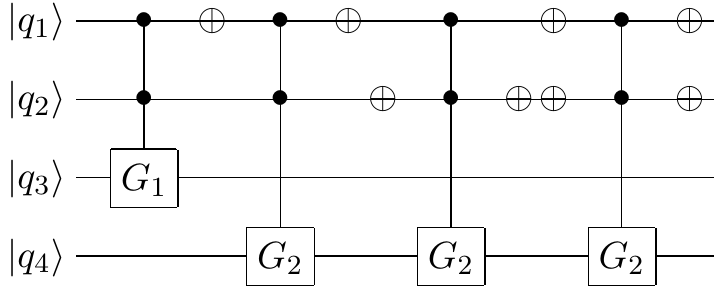}
}\\
\hline
\end{tabular}
 \caption{Example of quantum conditional operation with ``and'' operator.}
 \label{fig:condition-and}
\end{figure}

\begin{figure}[htp]
 \centering
\begin{tabular}{|l|l|}
\hline
\multicolumn{2}{|c|}{Or} \\
\hline
pseudocode & {\qo} \\
\hline
\begin{lstlisting}[language=qpc, mathescape=true]{}
qbit $q_1, q_2, q_3, q_4$

if ($q_1$ or $q_2$) then
  $\mathcal{G}_1(q_3)$
else
  $\mathcal{G}_2(q_4)$
\end{lstlisting}
&
\begin{lstlisting}[language=octave]{}
q1=newregister(1);
q2=newregister(1);
q3=newregister(1);
q4=newregister(1);

qif(...
	qror(...
		qreq(qureg(q1),1),...
		qreq(qureg(q2),1) ),..
	{G1,qureg(q3)}, ...
	{G2,qureg(q4)})
\end{lstlisting}\\
\hline
\multicolumn{2}{|c|}{
Quantum Circuit
}\\
\hline
\multicolumn{2}{|c|}{
	\includegraphics[width=0.8\columnwidth]{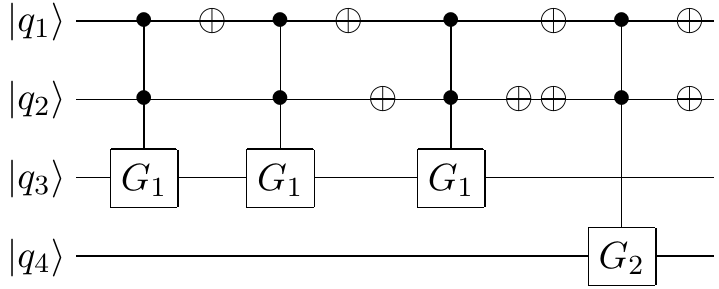}
}\\
\hline
\end{tabular}
 \caption{Example of quantum conditional operation with ``or'' operator.}
 \label{fig:condition-or}
\end{figure}

An example of quantum conditional gate controlled by logical expression defined
on quantum registers is presented in Figures~\ref{fig:condition-and}
and~\ref{fig:condition-or}.

\subsubsection{Quantum pointers}
In analogy to the concept of pointers and indirect addressing in classical
programming, one may introduce quantum pointers. The idea is to use control
register to control on which of the target registers an operation should be
applied. 

\begin{figure}[htp]
 \centering
\begin{tabular}{|l|l|}
\hline
\multicolumn{2}{|c|}{Pointer} \\
\hline
pseudocode & {\qo} \\
\hline
\begin{lstlisting}[language=qpc,mathescape=true]{}
qreg[2] $q_1$
qnibble $q_2$

if(*$q_1$) then
  $\mathcal{G}(q_2)$
\end{lstlisting}
&
\begin{lstlisting}[language=octave]{}
q1=newregister(2);
q2=newregister(4);

qpointer(G,qureg(q1),qureg(q2))
\end{lstlisting}\\
\hline
\multicolumn{2}{|c|}{
Quantum Circuit
}\\
\hline
\multicolumn{2}{|c|}{
 \includegraphics[width=0.8\columnwidth]{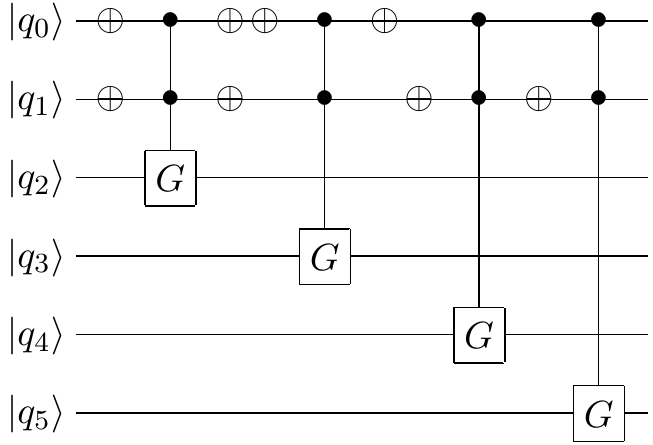}
}\\
\hline
\end{tabular}
 \caption{Example of simple quantum conditional operation controlled by quantum
 pointer.}
 \label{fig:quantum-pointer}
\end{figure}

Let us assume that one has the $n$-bit control register and a set of $2^n$-bit
target registers. The control register stores the address of target register to
which given unitary operation shall be applied. In order to visualise the use of
a~quantum pointer an example is shown in Figure~\ref{fig:quantum-pointer}.

Formally, a quantum pointer controlled by register $r_c$ with target $r_t$ is
constructed in the following way:
\begin{align}\label{equ:quantumif-pointer}
POINT_{r_t}(G_{[r_c]})= 
	\prod_{i\in\mathcal{P}(r_c)} \left(Not_{r_c\setminus i}G_{{\qvalueof{i}{r_c}|r_c}}Not_{r_c\setminus i}\right).
\end{align}

\section{Package \qo}\label{sec:qo}
Package {\qo} \cite{GM2004a, GM2004c, GM2005b} provides a quantum programming,
simulation and analysis language implemented as a~library of functions for GNU
Octave~\cite{Eaton2002}.

GNU Octave is computer algebra system (CAS) and high level programming language
designed primarily to perform numerical calculations. The basic data structure
in Octave is a matrix (integer, real or complex), therefore it is a natural
choice as a basis for the implementation of quantum programming language.

GNU Octave supports sparse matrices and distributed computing in shared and
distributed memory models. GNU Octave is a very flexible and easily extendible
tool. It is also free software and it can be used in a wide range of operating
systems.

\subsection{Design choices}
The main goal of the design of \qo\ is to provide a flexible and useful tool for
simulation of quantum information processing. Therefore it is based on GNU
Octave a high-level scientific programming language. This allows for a seamless
integration of very efficient matrix operations and numerical procedures with
the library of specialized functions provided by \qo. As GNU Octave is to large
extent compatible with Matlab, provided functions can be also used to simulate
and analyse quantum algorithms in Matlab.

One of the unique features of \qo\ is its ability to operate on both pure and
mixed quantum states. It allows to perform unitary as well as non-unitary
evolution represented by quantum channels.

Quantum gates can be constructed by the user in various ways: by calling
provided subroutines, by building their own subroutines, by using quantum
control structures. Additionally the user can build and use quantum channels or
use those already provided. Most of the \qo\ functions operate on quantum
registers and therefore the quantum operations build with their use are
re-allocable. 

A good illustration of those features is presented in the following
listing~\ref{lst:qft:qo} that contains the implementation of Quantum Fourier
transform in {\qo}. It can be compared to pseudo code version of the same
procedure listed in Listing~\ref{lst:qpc-example}. 
\begin{lstlisting}[language=octave, float=htpb, numbers=left, stepnumber=2,
numberstyle=\tiny, caption= Quantum Fourier transform in {\qo},
label=lst:qft:qo]{}
function ret=dft(gatesize)
  n=gatesize;
  cir=id(n);
  for i=[1:n]
    for j=[1:i-1]
      cir=circuit(cir,cphase(pi/(2^(i-j)),j,i,gatesize));
    endfor
    cir=circuit(cir,productgate(h,i,gatesize));
  endfor  
  ret=cir=circuit(cir, flip(n));
endfunction
\end{lstlisting}

Package {\qo} can operate on sparse and full matrices depending on users choice.
Sparse matrices need much less memory to store but operations on them may be
slower. Full matrices tend to consume huge memory space, but operations on them
are generally faster. In case of full matrices it should be possible to operate
on states of size up to ten qubits on a~contemporary workstation. Sparse
matrices should allow to simulate the quantum systems of up to 20 qubits.

Although \qo\ is not, strictly speaking, a~programming language ready to program
real quantum devices, with some effort it can be transformed in such a way that
it would be able to compile a high level programs to some sort of quantum
assembler. One should note that broad range of functions allowing the analysis
of states is implemented in the package. Those function are described in the
following section.

\subsection{Description}
Package {\qo} is designed to allow the user to operate on different levels of
abstraction. User can prepare complex gates and quantum channels from basic
primitives such as single qubit rotations, controlled gates, single qubit
channels. Most of the functions that form this library operate on quantum
registers, which makes the preparation of quantum gates and channels very
``natural''. The library is implemented in such a~way that depending of user's
choice it may operate on full or sparse matrices.

{\qo} can work in two modes: as a library or as a programming
language/simulator. Library mode is the default. To switch to language/simulator
mode one has call {\bt quantum\_octave\_init()} function. In case of the second
mode {\qo} allocates and manages an~internal quantum state and maintains the
quantum registers. Such functions as {\bt evolve()}, {\bt applychannel()}, {\bt
measure\-comp\-basis()} operate directly on the internal state. Listings of
Deustch's \ref{lst:deutsch} and Grover's \ref{lst:grover} algorithms show use of
lan\-gu\-age/si\-mu\-la\-tion mode.

\subsubsection{Convention}
Following conventions are used in {\qo}.
\begin{itemize}
	\item[--] \emph{Quantum register} is horizontal integer vector containing
	 indices of qubits starting from one.
	\item[--] \emph{Ket} is vertical complex vector. 
	\item[--] \emph{Bra} is horizontal complex vector.
	\item[--] \emph{Density matrix} is complex square matrix always of dimension
	 $n$-th power of two by $n$-th power of two.
	\item[--] \emph{Binary string} is 0,1-horizontal vector, that encodes a~binary
	  number. Order of bits is from MSB to LSB.
	\item[--] \emph{Size} of the gate or channel is always given in terms of a
	  number of qubits it acts on. If size is written in square brackets it means
	  that it can be omitted if the gate or channel acts on the whole system and
	  {\qo} was initialised.
\end{itemize}

\subsubsection{Quantum gates}
Package {\qo} supplies set of elementary gates known in quantum computation.
\begin{itemize}
\renewcommand{\labelitemi}{--}
	\item {\bt sx}, {\bt sy}, {\bt sz} -- return one-qubit Pauli operators {\bt
	  sx} -- $\sigma_x$, {\bt sy} -- $\sigma_y$, {\bt sz} -- $\sigma_z$.
	\item {\bt id(n)} -- returns identity matrix: $\1_n$.
	\item {\bt roty(a)}, {\bt rotz(a)}, {\bt rotx(a)} -- return rotation matrix
	  by angle {\bt a} around appropriate axis.
	\item {\bt qft(n)} -- returns quantum Fourier transform on {\bt n} qubits.
	\item {\bt swap(size, qubits)} -- returns swap gate of a given {\bt size}
	  that swaps {\bt qubits} given as two-element vector.
	\item {\bt qubitpermutation(permutation)} -- returns unitary gate that
	  performs gi\-ven {\bt permutation}.
	\item {\bt h} -- returns one-qubit Hadamard gate.
	\item {\bt phase(p0,p1)} -- returns one-qubit phase gate, with $p0, p1$ phase parameters.
\end{itemize}

\subsubsection{Basic functions}
Following functions are essential to prepare a~quantum state and to implement
a~quantum algorithm, protocol or game.
\begin{itemize}
\renewcommand{\labelitemi}{--}
	\item {\bt ket(binvec)} -- returns ket for given binary string.
	\item {\bt ketn(int,size)} -- returns a ket of size $2^{size}$ for given
	  {\bt int}eger number.
	\item {\bt state(pure\_state)} -- returns density matrix for a~given ket.
	\item {\bt mixstates(a1,mixed\_state1,[a2,\\ mixed\_state2,\ldots])} -- returns
	  con\-vex\\ combination of density matrices with coefficients {\bt a1, a2,
	  \ldots}.
	\item {\bt productgate(gate,targetreg[,size])} -- returns a~controlled gate
	  of a~gi\-ven {\bt size} that applies given {\bt gate} on {\bt target
	  reg}\-ister. See Eq.~\ref{equ:productgage}.
	\item {\bt controlledgate(gate,controlreg,\\ targetreg[,size])} -- returns a
	  controlled gate of given {\bt size} that applies {\bt gate} on specified
	  {\bt target reg}\-ister and is controlled by {\bt control reg}\-ister. See
	  Eq.~\ref{equ:controlledgate}.
\end{itemize}

\subsubsection{Quantum conditional operations}
The functions listed below implement quantum conditional operations, quantum
expressions and pointers. They are useful to simplify the implementation.
\begin{itemize}
\renewcommand{\labelitemi}{--}
	\item {\bt qif(expression,ifpart,elsepart,\\ size)} -- returns quantum gate of
	  given {\bt size}, controlled by {\bt expression} that applies {\bt ifpart}
	  if expression is true and {\bt elsepart} if expression is false. {\bt
	  ifpart} and {\bt elsepart} are cellarrays in the form: {\bt \{gate,
	  target\_register\}}. See Eq.~\ref{equ:quantumif}.
	\item {\bt qreq(register,integer)} -- returns expression: $[${\bt
	  register}$]$ equals {\bt integer}. See Eq.~\ref{equ:quantumif-relation}.
	\item {\bt qrne(register,integer)} -- returns expression: $[${\bt
	  register}$]$ not equal {\bt in\-teg\-er}. 
	\item {\bt qrge(register,integer)} -- returns expression: $[${\bt
	  register}$]$ is greater or equal to {\bt integer}.
	\item {\bt qrgt(register,integer)} -- returns expression: $[${\bt
	  register}$]$ is greater than {\bt integer}.
	\item {\bt qrle(register,integer)} -- returns expression: $[${\bt
	  register}$]$ is lesser or equal to {\bt integer}.
	\item {\bt qrlt(register,integer)} -- returns expression: $[${\bt
	  register}$]$ is lesser than {\bt integer}.
	\item {\bt qrin(register,set)} -- returns expression: $[${\bt register}$]$
	  is in {\bt set}.
	\item {\bt qror(expr1,expr2)} -- returns logical or on expressions {\bt
	  expr1} and {\bt expr2}. See Eq.~\ref{equ:quantumif-expression}.
	\item {\bt qrand(expr1,expr2)} -- returns logical and on expressions {\bt
	  expr1} and {\bt expr2}. 
    \item {\bt qpointer(gate,contrregister,\\ targteregister[,size])} -- returns
      quan\-tum gate of given {\bt size}, controlled by {\bt controll register}
      that applies {\bt gate} on {\bt target register}. See
      Eq.~\ref{equ:quantumif-pointer}.
\end{itemize}

\subsubsection{Evolution, channels and measurement}
The following group of functions allows to control the evolution of quantum
states and introduces the application of channels and measurement.

\begin{itemize}
\renewcommand{\labelitemi}{--}
 \item {\bt evolve(evolution[,state])} -- applies unitary {\bt evolution} to~the
   {\bt state}, returns the result of the evolution. See
   Eq.~\ref{equ:unitaryevolution}.
 \item {\bt channel(name,p)} -- returns Kraus operators acting on one qubit,
   para\-me\-tris\-ed by {\bt p} allowed {\bt name}s are: {\bt "depolarizing"},
   {\bt "amplitudedamping"}, \mbox{\bt "phasedamping"}, {\bt "bitflip"}, {\bt
   "phaseflip"} and {\bt "bitphaseflip"}. 
 \item {\bt localchannel(kraus, targetreg[, chsize])} -- returns a channel being
   the extension of defined by {\bt kraus} operators channel, acting on {\bt
   target reg}ister. 
 \item {\bt applychannel(elements[,state])} -- applies on~the~{\bt state} non
   unitary evolution defined by set of Kraus operators ({\bt elements}), returns
   the result of~the~evolution. See Eq.~\ref{equ:krausevolution}.
 \item {\bt ptrace(state, targetreg)} -- returns reduced density matrix for the
   {\bt state} with {\bt target reg}ister traced out. See
   Eq.~\ref{equ:traceevolution}.
 \item {\bt circuit(gate[, gate])} -- returns circuit composed of the sequence
   of {\bt ga\-te}s.
 \item {\bt measurecompbasis([state])} -- returns the probability distribution
   of the $\sigma_z$ measurement on the given {\bt state}.
 \item {\bt isunitary(gate)} -- returns {\bt true} if the {\bt gate} is unitary,
   otherwise returns {\bt false}.
 \item {\bt ischannel(operators)} -- returns {\bt true} if the {\bt operators}
   form valid quantum channel, otherwise returns {\bt false}.
 \item {\bt collapse(distribution)} -- chooses and returns a~basis state at
   random according to {\bt distribution}.
\end{itemize}

\subsubsection{Computation and control}
Following functions allows to control the quantum heap and configure the
behaviour of the library.
\begin{itemize}
\renewcommand{\labelitemi}{--}
 \item {\bt quantum\_octave\_init()} -- initialises the simulated system,
   creates quantum state with zero qubits allocated and empty list of registers.
 \item {\bt set\_quantum\_octave\_sparse([true | false])} -- switches on or off
   use of sparse matrices by all {\qo} functions.
 \item {\bt newregister(size)} -- creates new register of given {\bt size},
   allocates qubits on quantum heap, returns register id.
 \item {\bt clearregister(regid)} -- removes {\bt regid} register from quantum
   heap. Tra\-ces out appropriate qubits from the internal state.
 \item {\bt qureg(regid)} -- returns quantum register to~which {\bt regid}
   points.
 \item {\bt getstate()} -- returns the internal quantum state.
\end{itemize}

\subsubsection{Well known states}
Some of the states commonly used in quantum algorithms are implemented in the
library as separate functions.
\begin{itemize}
\renewcommand{\labelitemi}{--}
\item {\bt ghz(n)} -- returns Greenberger-Horne-Zeilinger state for {\bt n}
  qubits: $\frac{1}{\sqrt{2}}(\ket{0}^{\otimes n}+\ket{1})^{\otimes n}$.
\item {\bt phip} -- returns Bell $\ket{\Phi^+}$ state:
  $\frac{1}{\sqrt{2}}(\ket{00}+\ket{11})$.
\item {\bt phim} -- returns Bell $\ket{\Phi^-}$ state:
  $\frac{1}{\sqrt{2}}(\ket{00}-\ket{11})$.
\item {\bt psip} -- returns Bell $\ket{\Psi^+}$ state:
  $\frac{1}{\sqrt{2}}(\ket{01}+\ket{10})$.
\item {\bt psim} -- returns Bell $\ket{\Psi^-}$ state:
  $\frac{1}{\sqrt{2}}(\ket{01}-\ket{10})$.
\item {\bt maximallymixed(n)} -- return density matrix maximally mixed state:
  $\frac{1}{n}\1_n$.
\item {\bt wernersinglet(a)} -- returns 2-qubit Werner state:
  $a(\ket{00}-\ket{11})(\bra{00}-\bra{11}) + (1-a)\frac{\1}{4}$.
\end{itemize}

\subsubsection{Analysis}
Package {\qo} provides standard functions for analysis of quantum states, widely
used in quantum information literature. Among them the most important are:
\begin{itemize}
\renewcommand{\labelitemi}{--}
\item {\bt negativity(state, qubits)} -- computes negativity of the {\bt state}
  in respect to {\bt qu\-bits}.
\item {\bt entropy(state)} -- computes Von Neuman entropy of the {\bt state}.
\item {\bt concurrence(state)} -- computes concurrence of the {\bt state}.
\item {\bt fidelity(rho, sigma)} -- computes fidelity between density matrices
  {\bt rho} and {\bt sigma}.
\item {\bt fidelitypuremixed(psi, rho)} -- computes fidelity between ket {\bt
  psi} and density matrix {\bt sigma}.
\item {\bt tracenorm(state)} -- computes trace norm of the {\bt state}.
\item {\bt partialtranspose(state, targetreg)} -- returns matrix being partial
  tran\-spo\-si\-tion of {\bt state} matrix in~regard to~{\bt target reg}ister.
\end{itemize}

The next section presents the applications of {\qo} and various programming
techniques for solutions of quantum programming problems.

\section{Examples and applications}\label{sec:apps}
In what follows the applications of {\qo} and various high-level programming
techniques are discussed . It is shown how quantum processes, such as algorithms
may be implemented, simulated and analysed with this tool.

\subsection{Deutsch's problem}
One of the simplest quantum algorithms is Deutsch's algorithm. Although it may
seem trivial, this algorithm shows two very important features of quantum
computation. The first one is taht by taking the advantage of a superposition
one can compute any binary function for all its arguments in one step. The
second features is is that it is only possible to~retrieve information about
property of a~function and not on its values.

Let us assume that we have a black box that is usually called the oracle. This
box computes a function $f:\{0,1\}\rightarrow\{0,1\}$. We do not know if that
function is constant $f(0)=f(1)$ or injective $f(0)=\overline{f(1)}$. In
classical case we have to ask the oracle twice to check which kind the function
$f$ is. But in quantum case it is possible to solve this problem asking the
oracle only once.

The algorithm goes as follows:
\begin{enumerate}
\item Prepare the state: $\ket{\Psi}=\ket{0}\otimes\ket{1},$.
\item Apply the Hadamard $H^{\otimes 2}$ gate on the state $\ket{\Psi}$, you will get
\begin{equation}
\ket{\Psi_1}=\frac{\ket{0}+\ket{1}}{\sqrt{2}}\otimes\frac{\ket{0}-\ket{1}}{\sqrt{2}}.
\end{equation}
\item Apply the gate
$
U_f:\ket{x}\otimes\ket{y}\rightarrow\ket{x}\otimes\ket{f(x)\oplus y}
$
on the state $\ket{\Psi_1}$; you will get:
\begin{equation}
\ket{\Psi_2}=\left\{
\begin{array}{ll}
    \pm\frac{\ket{0}+\ket{1}}{\sqrt{2}}\otimes\frac{\ket{0}-\ket{1}}{\sqrt{2}}&\text{for constant $f$},\\
    \pm\frac{\ket{0}-\ket{1}}{\sqrt{2}}\otimes\frac{\ket{0}-\ket{1}}{\sqrt{2}}&\text{for injective $f$}.\\
\end{array}\right.
\end{equation}
\item Apply $H\otimes \1$ on the state $\ket{\Psi_2}$; you will get:
\begin{equation}
\ket{\Psi_3}=\left\{
\begin{array}{ll}
    \pm\ket{0}\otimes\frac{\ket{0}-\ket{1}}{\sqrt{2}}&\text{for constant $f$},\\
    \pm\ket{1}\otimes\frac{\ket{0}-\ket{1}}{\sqrt{2}}&\text{for injective $f$}.\\
\end{array}\right.
\end{equation}
\item Measure state of the first qubit, you will get $\ket{0}$ in case of
constant function, $\ket{1}$ for injective function.
\end{enumerate}

\begin{figure}[ht]
    \begin{center}
        \includegraphics[width=0.75\columnwidth]{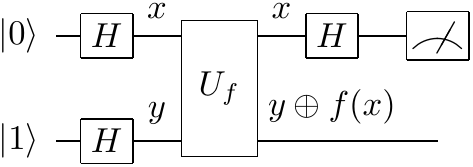}
    \end{center}
    \caption{Deutsch's algorithm}
    \label{fig:deustch}
\end{figure}

Quantum circuit representation of Deutsch's algorithm is presented in
Figure~\ref{fig:deustch}. The $U_f$ gate provide a reversible implementation of
function $f$ and the symbol \includegraphics[scale=.2]{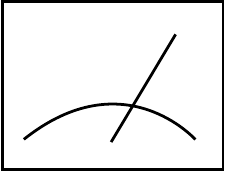} denotes
a measurement.

The implementation of Deutsch's algorithm presented in listing~\ref{lst:deutsch}
is an introductory example of application of \qo\ for simulation of a~quantum
algorithm with all basic steps of computation: initialization of the quantum
computer, unitary evolution and measurement.

Below we describe simulation steps (compare with circuit in
Figure~\ref{fig:deustch}). Numbers on the left refer to the line numbers in
Listing~\ref{lst:deutsch}.
\begin{description}
 \item[5]: initialisation of the simulator,
 \item[7, 8]: allocation of registers,
 \item[10-17]: definition of all four possible oracles,
 \item[19]: application of $Not$ on second qubit,
 \item[20]: application of $H\otimes H$,
 \item[21]: application of the oracle,
 \item[22]: application of $H\otimes \1$,
 \item[24]: tracing out of second register,
 \item[26]: return the probability distribution of the measurement outcome. 
\end{description}

\begin{lstlisting}[language=octave, numbers=left, stepnumber=2, numberstyle=\tiny, 
caption=Deutsch algorithm in {\qo}, label=lst:deutsch]{}
# input: identifier of the function
# output: state after execution of Deutsch's algorithm
function ret = deutsch(num)
	# initialize the simulation
	quantum_octave_init();
	# declare and allocate registers	
	r1=newregister(1);
	r2=newregister(1);
	# declare functions
	f{1}=id(2);
	f{2}=productgate(sx,qureg(r2));
	f{3}=qif(qreq(qureg(r1),1),...
			{sx,qureg(r2)},...
			{id,qureg(r2)});
	f{4}=qif(qreq(qureg(r1),0),...
			{sx,qureg(r2)},...
			{id,qureg(r2)});
	# do the algorithm
	evolve(productgate(sx,qureg(r2)));
	evolve(productgate(h,[qureg(r1),qureg(r2)]));	
	evolve(f{num});	
	evolve(productgate(h,qureg(r1)));	
	# throw away second register
  clearregister(qureg(r2));
	# return the outcome
	ret=measurecompbasis();
endfunction
\end{lstlisting}

\subsection{Grover's algorithm}
To illustrate more advanced usage of the presented concepts we use the quantum
algorithm for searching a unordered database. The algorithms was proposed by
Grover~\cite{Grover1996, Grover1997, Grover1998} and its detailed description
and analysis can be found in \cite{Bugajski2001, lomonaco02grover}. Here we
present the implementation of Grover's algorithm which presents the features of
\qo\ related to the observation of quantum errors. We show the propagation of
initial errors during the execution of the algorithm.

Grover's search algorithm is one of the most important quantum algorithms. This
is especially true since many algorithmic problems can be reduced to exhaustive
search. However, like in the case of any quantum procedure, the efficiency of
the algorithm depends on the ability to avoid errors during the procedure.
Thus, it is important how quantum errors affect the executions of the algorithm.

\subsubsection{Statement of the problem}
Let $X$ be a~set and let $f: X\rightarrow\{0,1\}$, such that
\begin{equation}
    f(x)=
    \left\{
    \begin{array}{l}
        1 \Leftrightarrow x= x_0\\
        0 \Leftrightarrow x\neq x_0
    \end{array}
    \right., x\in X,
\end{equation}
for some marked $x_0\in X$.

For the simplicity we assume that $X$ is a set of binary strings of length $n$.
Therefore $|X|=2^n$ and $f:\{0,1\}^n \rightarrow \{0,1\}$.

We can map the set $X$ to the set of states over $\HH^{\otimes n}$ in the
natural way as
\begin{equation}
x \leftrightarrow \ket{x}.
\end{equation}

The goal of the algorithm is to find the marked element. This is achieved by the
amplification of the appropriate amplitude~\cite{Bugajski2001,
lomonaco02grover}. 

\subsubsection{The algorithm}
The Grover's algorithm is composed of two main procedures: the oracle and the
diffusion.

\paragraph{Oracle}
By oracle we understand a function that marks one defined element. In the case
of this algorithm, the marking of the element is done by negation of the
amplitude of the state that we search for.

With the use of elementary quantum gates the oracle can be constructed using
an ancilla $\ket{q}$ in the following way:
\begin{equation}
\label{equ:oracle1}
   O\ket{x}\ket{q}=\ket{x}\ket{q\otimes f(x)}.
\end{equation}
If the register $\ket{q}$ is prepared in the state: 
\begin{equation}
    \ket{q}=H\ket{1}=\frac{\ket{0}-\ket{1}}{\sqrt{2}},
\end{equation}
then by substitution, equation \ref{equ:oracle1} is re-transformed to:
\begin{equation}
    O\ket{x}\frac{\ket{0}-\ket{1}}{\sqrt{2}}=(-1)^{f(x)}\ket{x}\frac{\ket{0}-\ket{1}}{\sqrt{2}},
\end{equation}
and by tracing out the ancilla we get:
\begin{equation}
    O\ket{x}=-(-1)^{f(x)}\ket{x}.
\end{equation}
Thus the oracle marks a given state by inverting its amplitude.

\paragraph{Diffusion}\label{sec:diffusion}
The operator $D$ rotates any state around the state
\begin{equation}
\ket{\psi}= \frac{1}{\sqrt {2^n}}\sum\limits_{x = 0}^{2^n-1}\ket{x},
\end{equation}
$D$ may written in the following form:
\begin{equation}
D=-H^{\otimes n}(2\ketbra{0}{0}-\1)H^{\otimes
n}=2\ketbra{\psi}{\psi}-\1.
\end{equation}

\paragraph{Grover iteration}
The first step of the algorithm is to apply Hadamard gate $H^{\otimes n}$ on all
the qubits. Then we apply gate $G=DO$ several times.

\paragraph{Number of iterations}
Application of diffusion operator on the base state $\ket{n}$ gives
\begin{equation}
-H^{\otimes n}I_{0}H^{\otimes n}\ket{n}=-\ket{x_0}+\frac{2}{N}\sum_y\ket{y}.
\end{equation}
Application of this operator on any state gives
\begin{eqnarray*}
D\ket{x}&=&\sum_{x}\alpha_{x}( -\ket{x} +\frac{2}{N}y\sum_y\ket{y}) \\
&=&\sum_{x}( -\alpha_{x}+2s) \ket{x},
\end{eqnarray*}
where
\begin{equation}
s=\frac 1N\sum_{\mathbf{x}}\alpha _{\mathbf{x}}
\end{equation}
is arithmetic mean of coefficients $\alpha_x,\
x=0,\ldots,2^n-1$.

$k$-fold application of Grover's iteration $G$ on initial state $\ket{s}$ leads
to \cite{Bugajski2001}:
\begin{equation}\label{equ:grover1}
    G^k\ket{s} =\alpha_k\sum_{x\neq x_0}\ket{x}+\beta_k\ket{x_0},
\end{equation}
with real coefficients:
\begin{equation}\label{equ:grover2}
    \alpha _k=
    \frac{1}{\sqrt{N-1}} \cos\left(2k+1\right)
    \theta ,\quad \beta _k=\sin\left(2k+1\right) \theta,
\end{equation}
where $\theta $ is an angle that fulfils the relation:
\begin{equation}
    \sin(\theta) =\frac{1}{\sqrt{N}}.
\end{equation}
Therefore the coefficients $\alpha _k,\beta _k$ are periodic functions of $k$.
After several iterations amplitude of $\beta_k$ rises and others drop. The
influence of the marked state $\ket{x_0}$ on the state of the register is that
the initial state $\ket{s}$ evolves towards the marked state.

The $\beta_k$ attains its maximum after approximately $\frac{\pi}{4}\sqrt{N}$
steps. Then it begins to fall. Thus, the number of steps needed to transfer the
initial state towards the marked state is of $O(\sqrt{n})$. In the classical
case the number of steps is of $O(n)$.

\paragraph{Measurement} 
The last step of the Grover's algorithm is the measurement. Probability of
obtaining of the proper result is $|\beta_k|^2$.

\begin{figure}[ht]
    \begin{center}
        \includegraphics[width=\columnwidth]{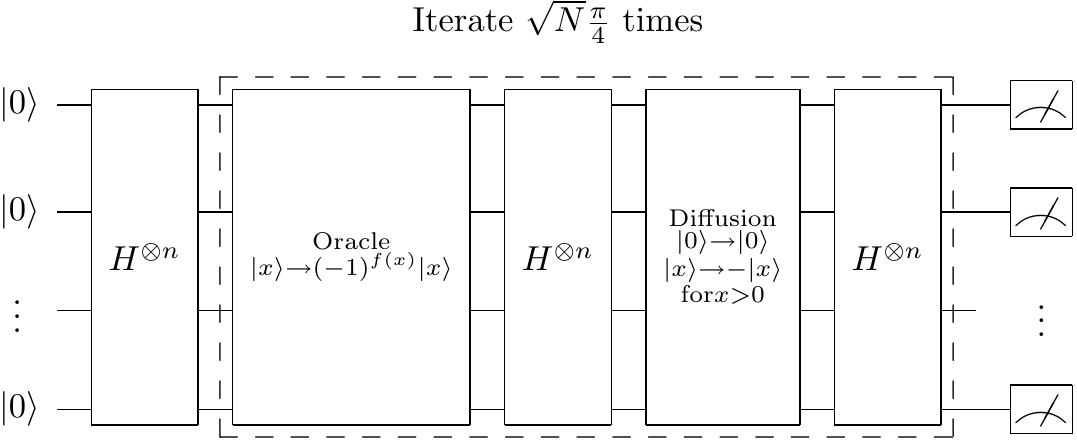}
    \end{center}
    \caption{The circuit for Grover's algorithm.}
    \label{fig:grover}
\end{figure}

\subsubsection{Graphical interpretation}
There a~exists very nice graphical interpretation of Grover's algorithm.

Let $\ket{\alpha}$ denotes the sum of states orthogonal to the state we are
searching for $\ket{x_0}$
\begin{equation}
    \ket{\alpha}=\frac{1}{\sqrt{2^n-1}}\sum_{x\neq x_0}\ket{x},
\end{equation}
and for consistence we will write $\ket{\beta}=\ket{x_0}$. Then, on the plane
spanned by $\ket{\alpha}$ and~$\ket{\beta}$, we can observe of evolution of the
state vector.

\begin{figure}[ht]
    \begin{center}
        \includegraphics[scale=0.4]{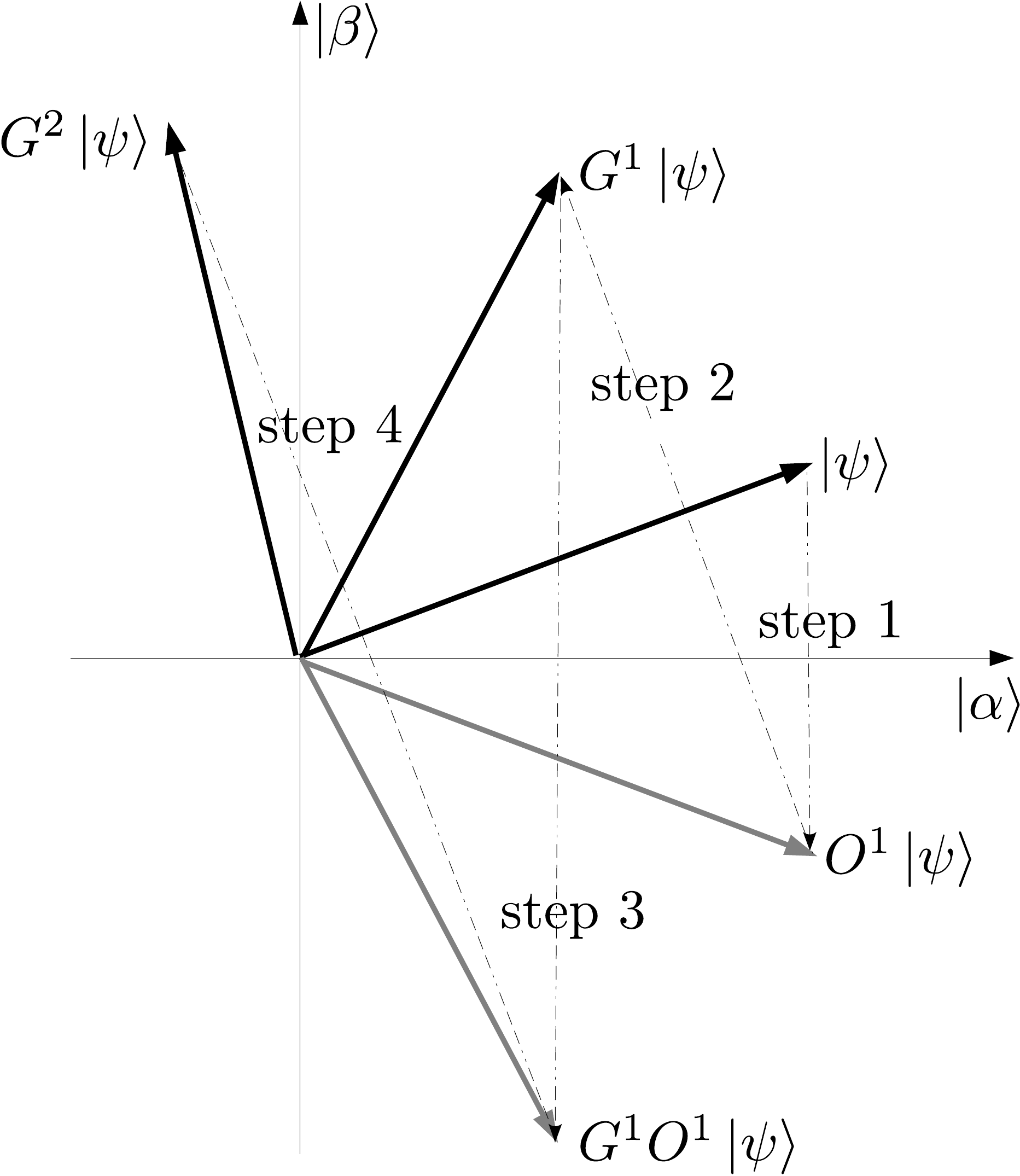}
    \end{center}
    \caption[Visualisation of Grover's algorithm.]{Visualisation of Grover's
    algorithm \cite{NC2000}. Projection on plane spanned by $\ket{\alpha}$ and
    $\ket{\beta}$. Vector $\ket{\psi}$ is flat superposition of all the possible
    states.}
    \label{fig:grover:wiz}
\end{figure}

By putting values from equation \ref{equ:grover2} into equation
\ref{equ:grover1} we get following relation:
\begin{equation}\label{equ:grover3}
    G^k\ket{s} = \cos((2k+1)\theta)\ket{\alpha}+
    \sin((2k+1)\theta)\ket{\beta}.
\end{equation}
Exemplar behaviour of this equation for $2^3$ states is presented in Figure~\ref{fig:grover:wiz}.

\subsubsection{Implementation}
Listing~\ref{lst:grover} presents the implementation of function {\bt
grover}. We will apply quantum noise at the end of each Grover iteration and
observe its influence on its efficiency.

To simulate this behaviour we will insert the code from
Listing~\ref{lst:grover:noise} after line 21 of the implementation.

\begin{lstlisting}[language=octave, numbers=left, stepnumber=2, numberstyle=\tiny, caption=Grover's algorithm in \qo\ , label=lst:grover]{}
# function implementing Grover's algorithm
# input: number we are looking for, size of the system
# output: probability distribution after 
#					execution of the algorithm
function ret = grover(num, S)
	# initialize the simulation
	quantum_octave_init();
	# allocate register
	r1=newregister(S);
	# number of elements
	N=2^length(qureg(r1))
	# calculate number of iterations
	k = floor((pi/4)*sqrt(N));
	# prepare the system in flat superposition of base states
	evolve(productgate(h,qureg(r1)));
	# Grover iterations
	for i = 1:k
		# ask the oracle
		evolve(oracle(num, qureg(r1)));
		# diffuse
		evolve(diffuse(qureg(r1)));
	endfor
	# return probability distribution of base states
 	ret=measurecompbasis();
endfunction

# function implementing oracle 
# input: number to mark, size of the system
# output: gate implementing oracle of the size 2^l
function ret = oracle (num, register)
	l=length(register);
	ret = id(l);
	ret(num+1,num+1) = -1;
endfunction

# function implementing oracle 
# input: register on which implement diffusion
# output: diffusion gate of the size 2^l
function ret = diffuse(register)
	l=length(register);
	ret = circuit(...
					productgate(h,register,l),...
					(2*ketn(0,l)*bran(0,l) - id(l)),...
					productgate(h,register,l)...
				);
endfunction
\end{lstlisting}

\begin{lstlisting}[language=octave, numbers=left, stepnumber=2,
numberstyle=\tiny, caption=Adding noise to Grover's algorithm,
label=lst:grover:noise]{}
applychannel(
    localchannel(
        channel(channelname,p),qureg(r1)
    )
);
\end{lstlisting}

\begin{figure}[htpb!]
\begin{center}
\includegraphics[width=\columnwidth]{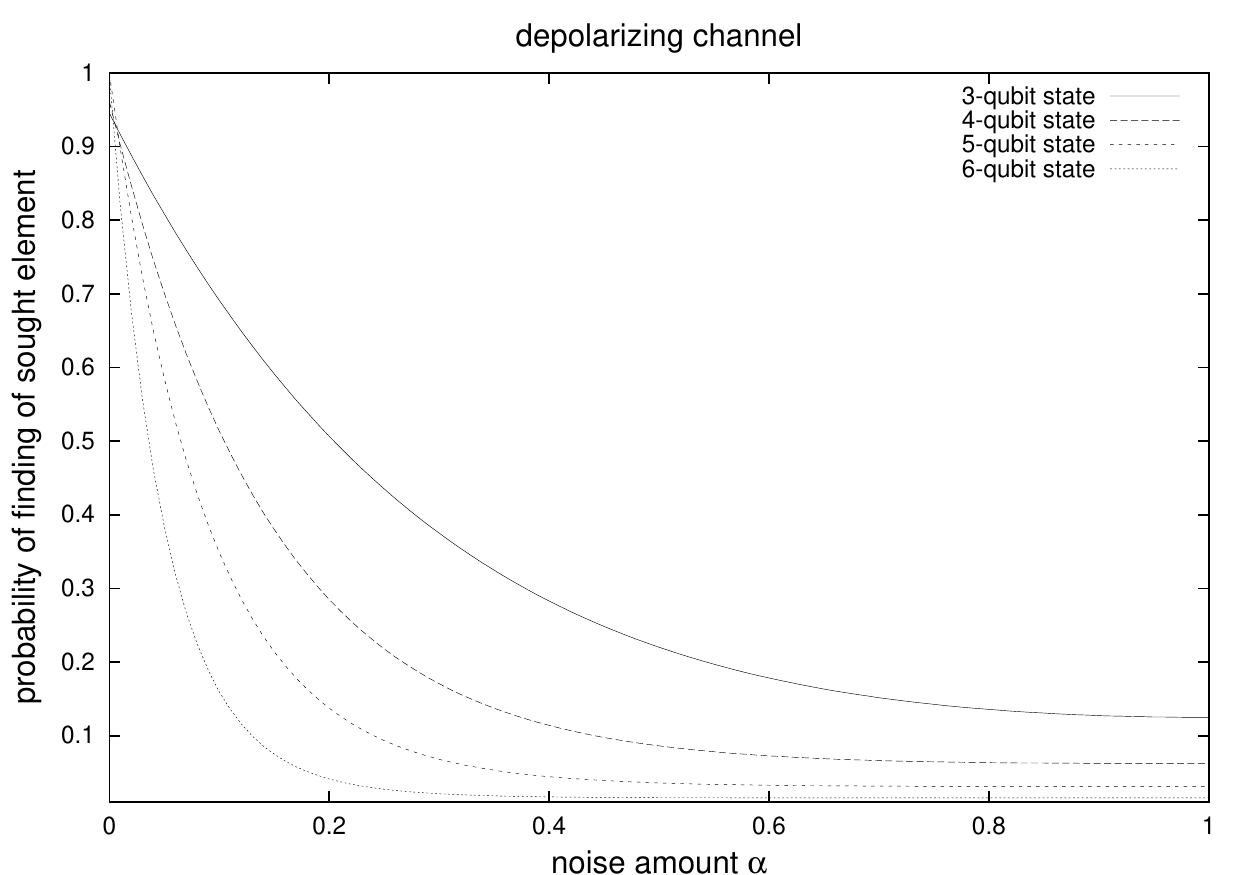}
\caption[Grover's algorithm under the depolarizing channel.]{Influence of
depolarizing channel parametrized by single real number $\alpha$ on probability
of successful finding of sought element in Grover's algorithm implemented with
3, 4, 5 and 6 qubits.}
\label{fig:grover:depolarizing_channel}
\end{center}
\end{figure}

\paragraph{Simulation results}
The results of the simulation of noisy Grover's algorithm acting on system of
size from three to six qubits when system is affected by noise modelled with
depolarizing channel are shown in Figure~\ref{fig:grover:depolarizing_channel}.
One may observe that rate of successful application of the algorithm drops
quickly with raising amount of noise. This effect is more significant for larger
systems. This result clearly indicates that it is not possible to successfully
implement Grover's algorithm in presence of large amounts of noise if no error
correction scheme is applied. 

\section{Summary}\label{sec:summary}
We have introduced an original solution to the problem of simulation of quantum
processes. This solution is provided by \qo, a~library that is build upon GNU
Octave high level programming language, which provides high-level quantum
programming structures. 

Although strictly speaking, \qo\ is not a~programming language but a~library,
together with GNU Octave, it is very convenient and flexible tool. Programs
written in quantum programming languages, such as QCL, can be easily rewritten
using this library, thanks to the use of quantum memory, registers and routines.
Scalable programs can be easily implemented in \qo, so the programmer does not
have to think about details of the implementation.

\section*{Acknowledgements}
We acknowledge the financial support by the Polish Ministry of Science and
Higher Education under the grant number N519 012 31/1957 and by the Polish
research network LFPPI. The numerical calculations presented in this work were
performed on the \texttt{Leming} server of The Institute of Theoretical and
Applied Informatics of the Polish Academy of Sciences.

Package \qo\ is distributed as free software and it can be downloaded from the
project web-page \cite{qo}.



\begin{thebibliography}{10}
\providecommand{\url}[1]{#1}
\csname url@samestyle\endcsname
\providecommand{\newblock}{\relax}
\providecommand{\bibinfo}[2]{#2}
\providecommand{\BIBentrySTDinterwordspacing}{\spaceskip=0pt\relax}
\providecommand{\BIBentryALTinterwordstretchfactor}{4}
\providecommand{\BIBentryALTinterwordspacing}{\spaceskip=\fontdimen2\font plus
\BIBentryALTinterwordstretchfactor\fontdimen3\font minus
  \fontdimen4\font\relax}
\providecommand{\BIBforeignlanguage}[2]{{%
\expandafter\ifx\csname l@#1\endcsname\relax
\typeout{** WARNING: IEEEtran.bst: No hyphenation pattern has been}%
\typeout{** loaded for the language `#1'. Using the pattern for}%
\typeout{** the default language instead.}%
\else
\language=\csname l@#1\endcsname
\fi
#2}}
\providecommand{\BIBdecl}{\relax}
\BIBdecl

\bibitem{NC2000}
M.~A. Nielsen and I.~L. Chuang, \emph{{Quantum Computation and Quantum
  Information}}.\hskip 1em plus 0.5em minus 0.4em\relax Cambridge University
  Press, 2000.

\bibitem{hirvensalo}
M.~Hirvensalo, \emph{Quantum computing}.\hskip 1em plus 0.5em minus 0.4em\relax
  Springer, 2001.

\bibitem{BKW2001a}
S.~Bugajski, J.~Klamka, and S.~Wegrzyn, ``Foudations of quantum computing.
  {Part I},'' \emph{Archiwum Informatyki Teoretycznej i Stosowanej}, vol.~13,
  no.~1, pp. 97--142, 2001.

\bibitem{BKW2001b}
------, ``Foudations of quantum computing. {Part II},'' \emph{Archiwum
  Informatyki Teoretycznej i Stosowanej}, vol.~13, no.~1, pp. 137--149, 2001.

\bibitem{shor03why}
P.~W. Shor, ``Why haven't more quantum algorithms been found?'' \emph{Journal
  of the ACM}, vol.~50, no.~1, pp. 87--90, 2003.

\bibitem{shor04progress}
------, ``Progress in quantum algorithms,'' \emph{Quantum Information
  Processing}, vol.~3, no. 1-5, 2004.

\bibitem{Deutsch1985}
D.~Deutsch, ``Quantum theory, the {Church-Turing} principle and the universal
  quantum computer,'' \emph{Proc. Roy. Soc. Lond.}, vol. A 400, p.~97, 1985,
  \href{http://www.qubit.org/people/david/David.html}{http://www.qubit.org/peo%
ple/david/David.html}.

\bibitem{Deutsch1989}
------, ``Quantum computational networks,'' \emph{Proc. Roy. Soc. Lond.}, vol.
  A 425, p.~73, 1989.

\bibitem{bettelli03towards}
S.~Bettelli, L.~Serafini, and T.~Calarco, ``Toward an architecture for quantum
  programming,'' \emph{Eur. Phys. J. D}, vol.~25, no.~2, pp. 181--200, 2003.

\bibitem{Gudder2003}
S.~Gudder, ``Quantum computational logic,'' \emph{International Journal of
  Theoretical Physics}, vol.~1, no.~42, pp. 39--47, 2003.

\bibitem{Vantonder2004}
\BIBentryALTinterwordspacing
A.~{van Tonder}, ``A lambda calculus for quantum computation,'' \emph{SIAM
  J.COMPUT.}, vol.~33, p. 1109, 2004. [Online]. Available:
  \url{doi:10.1137/S0097539703432165}
\BIBentrySTDinterwordspacing

\bibitem{Moore2000275}
\BIBentryALTinterwordspacing
C.~Moore and J.~P. Crutchfield, ``Quantum automata and quantum grammars,''
  \emph{Theoretical Computer Science}, vol. 237, no. 1-2, pp. 275 -- 306, 2000.
  [Online]. Available:
  \url{http://www.sciencedirect.com/science/article/B6V1G-43FWYP1-H/2/201e1de6%
cd57eb5d309d61622092bd1d}
\BIBentrySTDinterwordspacing

\bibitem{bernstein97complexity}
\BIBentryALTinterwordspacing
E.~Bernstein and U.~Vazirani, ``Quantum complexity theory,'' \emph{{SIAM}
  Journal on Computing}, vol.~26, no.~5, pp. 1411--1473, 1997. [Online].
  Available: \url{http://www.cs.berkeley.edu/~vazirani}
\BIBentrySTDinterwordspacing

\bibitem{gay05quantum}
S.~Gay, ``Quantum programming languages: Survey and bibliography,''
  \emph{Bulletin of the European Association for Theoretical Computer Science},
  2005.

\bibitem{MiszczakPHD}
J.~A. Miszczak, ``{Probabilistic aspects of quantum programming languages},''
  Ph.D. dissertation, The Institute of Theoretical and Applied Informatics of
  the Polish Academy of Sciences, 2008.

\bibitem{Gay2008}
S.~Gay, ``Bibliography on quantum programming languages,'' 2007, web-page
  http://www.dcs.gla.ac.uk/\~{}simon/quantum/.

\bibitem{altenkirch05functional}
T.~Altenkirch and J.~Grattage, ``A functional quantum programming language,''
  in \emph{Proceedings. 20th Annual IEEE Symposium on Logic in Computer
  Science}.\hskip 1em plus 0.5em minus 0.4em\relax IEEE Computer Society, 2005,
  pp. 249--258.

\bibitem{knill96conventions}
E.~Knill, ``Conventions for quantum pseudocode,'' Los Alamos National
  Laboratory, Tech. Rep. LAUR-96-2724, 1996.

\bibitem{Oemer2003}
B.~Oemer, ``Structured quantum programming,'' Ph.D. dissertation, Technical
  University of Vienna, 2003.

\bibitem{cook73time}
S.~A. Cook and R.~A. Reckhow, ``Time-bounded random access machines,'' in
  \emph{Proceeedings of the forth Annual {ACM} Symposium on Theory of
  Computing}, 1973, pp. 73--80.

\bibitem{papadimitriou}
C.~H. Papadimitriou, \emph{Computational complexity}.\hskip 1em plus 0.5em
  minus 0.4em\relax Addison-Wesley Publishing Company, 1994.

\bibitem{shepherdson63computability}
J.~C. Shepherdson and H.~E. Strugis, ``Computability of recursive functions,''
  \emph{Journal of the {ACM}}, vol.~10, no.~2, pp. 217--255, April 1963.

\bibitem{cleve96schumacher}
R.~Cleve and D.~P. DiVincenzo, ``Schumacher's quantum data compression as a
  quantum computation,'' \emph{Phys. Rev. A}, vol.~54, no.~4, pp. 2636--2650,
  Oct 1996.

\bibitem{cormen}
T.~H. Cormen, C.~E. Leiserson, R.~L. Rivest, and C.~Stein, \emph{Introduction
  to Algorithms}, $2^{nd}$~ed.\hskip 1em plus 0.5em minus 0.4em\relax The MIT
  Press, 2001.

\bibitem{hopcrof03wprowadzenie}
J.~E. Hopcroft and J.~D. Ullman, \emph{Wprowadzenie do teorii automat\'ow,
  j\k{e}zyk\'ow i oblicze\'n}.\hskip 1em plus 0.5em minus 0.4em\relax
  Wydawnictwo Naukowe PWN, 2003.

\bibitem{bettelliPHD}
S.~Bettelli, ``Toward an architecture for quantum programming,'' Ph.D.
  dissertation, Universit\`a di {Trento}, February 2002.

\bibitem{GawronPHD}
P.~Gawron, ``High level programming in quantum computer science,'' Ph.D.
  dissertation, The Institute of Theoretical and Applied Informatics of the
  Polish Academy of Sciences, 2008.

\bibitem{Oemer2000}
B.~Oemer, ``Quantum programming in {QCL},'' Master's thesis, TU Viena, 2000.

\bibitem{GM2004a}
P.~Gawron and J.~A. Miszczak, ``Didactic tools for teaching quantum
  informatics,'' \emph{Annales UMCS Informatica AI}, vol.~1, no.~2, 2004.

\bibitem{GM2004c}
------, ``Simulations of quantum systems evolution with quantum-octave
  package,'' \emph{Annales UMCS Informatica AI}, vol.~1, no.~2, 2004.

\bibitem{GM2005b}
\BIBentryALTinterwordspacing
------, ``Numerical simulations of mixed states quantum computation,''
  \emph{Int. J. Quan. Inf.}, vol.~3, no.~1, pp. 195--199, 2005. [Online].
  Available: \url{doi:10.1142/S0219749905000748}
\BIBentrySTDinterwordspacing

\bibitem{Eaton2002}
J.~W. Eaton, \emph{{GNU Octave Manual}}.\hskip 1em plus 0.5em minus 0.4em\relax
  Network Theory Limited, 2002.

\bibitem{Grover1996}
L.~Grover, ``A fast quantum mechanical algorithm for database search,'' in
  \emph{Proc. 28th Annual ACM Symposium on the Theory of Computation}.\hskip
  1em plus 0.5em minus 0.4em\relax New York, NY: ACM Press, New York, 1996, pp.
  212--219.

\bibitem{Grover1997}
L.~K. Grover, ``Quantum mechanics helps in searching for a needle in a
  haystack,'' \emph{Phys. Rev. Lett.}, vol.~79, p. 325, 1997,
  \href{http://www.arxiv.org/abs/quant-ph/9706033}{arXiv:quant-ph/9706033}.

\bibitem{Grover1998}
------, ``A framework for fast quantum mechanical algorithms,'' in
  \emph{Proceedings of 30th Annual {ACM} Symposium on Theory of Computing
  {(STOC)}}, 1998, pp. 53--62,
  \href{http://www.arxiv.org/abs/quant-ph/9711043}{arXiv:quant-ph/9711043}.

\bibitem{Bugajski2001}
S.~Bugajski, ``Quantum search,'' \emph{Archiwum Informatyki Teoretycznej i
  Stosowanej}, vol. Tom 13, no. z. 2, pp. 143--150, 2001.

\bibitem{lomonaco02grover}
S.~J. Lomonaco, ``{Grover's} quantum search algorithm,'' \emph{Proceedings of
  Symposia in Applied Mathematics}, vol.~58, pp. 181--192, 2002.

\bibitem{qo}
``Project quantum-octave,'' web-page http://quantum-octave.sf.net/.

\end{thebibliography}
\end{document}